\documentclass[aps,prl,superscriptaddress,twocolumn,floatfix,showpacs]{revtex4}
\usepackage{amsmath}
\usepackage{graphicx}

\begin{document}

\title{Efficient mixed-force first-principles molecular dynamics}

\author{ Eduardo Anglada }
\affiliation{ Departamento de F\'{\i}sica de la Materia Condensada, C-III,
              Universidad Aut\'{o}noma de Madrid,
              E-28049 Madrid, Spain }

\author{ Javier Junquera }
\affiliation{ Departamento de F\'{\i}sica de la Materia Condensada, C-III,
              Universidad Aut\'{o}noma de Madrid,
              E-28049 Madrid, Spain }
\affiliation{ Institut de Physique, B\^atiment B5, Universit\'e de Li\`ege,
              B-4000 Sart-Tilman, Belgium }

\author{ Jos\'e M. Soler }
\affiliation{ Departamento de F\'{\i}sica de la Materia Condensada, C-III,
              Universidad Aut\'{o}noma de Madrid,
              E-28049 Madrid, Spain }

\date{\today}

\begin{abstract}
   We present an efficient method to mix well converged ab initio forces
with simpler and faster ones in molecular dynamics.
   While the cheap forces are evaluated every time step, the converged
ones correct the trajectory only every $n$ time steps.
   For convenience, both types of forces are calculated with the same 
basic scheme, using density functional theory, norm conserving 
pseudopotentials, and a basis set of numerical atomic orbitals.
   The cheap forces are evaluated with a short-range minimal basis set
and the non-selfconsistent Harris functional.
   Since these evaluations are hundreds of times faster than those of
the converged forces, they add a neglegible cost, and the boost in 
computational efficiency is approximately a factor $n$.
   Our results indicate that one can use values of $n$ of up to 10,
without affecting significantly the calculated structural and
dynamical magnitudes.
\end{abstract}



\maketitle

   Molecular dynamics (MD) is a fundamental tool in atomistic materials
simulation~\cite{Allen-Tildesley1987}.
   A majority of practicioners have used classical, semiempirical 
interatomic potentials.
   This is necessary for the large sizes and long times required to
simulate many processes of enormous scientific and technological interest, 
from materials deformation and fracture~\cite{Bulatov1998} to protein 
folding~\cite{Duan-Kollman1998}.
   A large effort has been devoted to develop interatomic potentials for
many types of systems~\cite{Voter1996}.
   However, the quantitative reliability of such potentials in situations 
of bond formation and breaking is highly questionable.
   In such cases, it is imperative to use the much more expensive 
ab initio MD methods~\cite{Car-Parrinello1985,Payne1992:RMP}, generally 
limited to a few hundred atoms and a few tens of picoseconds.
   Thus, it is essential to find methods that accelerate the integration
of the dynamical equations, thus allowing for longer simulations.

   In classical dynamics, one of such methods~\cite{Tuckerman1992} 
uses multiple time scales to integrate the equations of motion for  
systems with both fast and slow dynamical degrees of freedom.
   The same method can be used to compute separately the hard,
short-ranged forces from the soft, long-ranged ones.
   De Vita and Car~\cite{deVitaCar1998} have proposed to adapt 
`on the fly' the parameters of a classical potential using sporadic
or periodic evaluations of ab initio forces.
   In this work, drawing ideas of those previous works, we propose a 
simple method to speed up dramatically ab initio MD.
   In principle, it could be implemented by combining classical and 
ab initio forces.
   However, such an approach would still require to develop a suitable
classical force field for every new system with different interactions.
   Therefore, instead we take advantage of the fact that, while standard 
density functional forces require the simultaneous convergence of many 
parameters, much lower values of those parameters can still yield quite 
reasonable forces.
   Thus, by reducing drastically the size of the basis set, the 
Brillouin zone sampling, or the number of selfconsistency iterations, 
it is possible to reduce the computer time by enormous factors, 
and still obtain forces which are considerably more reliable than 
those of classical interatomic potentials.

   To test our scheme, we have chosen the SIESTA 
method~\cite{Ordejon1996,Soler2002}, which is specially well suited 
to span the range from `quick and dirty' calculations to fully 
converged ones.
   It uses density functional theory~\cite{Kohn-Sham1965} (DFT),
norm-conserving pseudopotentials~\cite{Troullier-Martins1991} 
and a basis set of numerical atomic orbitals of strictly finite 
range~\cite{Sankey-Niklewski1989,Anglada2002}.
   To calculate converged forces, we might typically use the generalized 
gradient approximation (GGA) to exchange and correlation (with spin
polarization if required), double-$\zeta$ polarized (DZP) basis orbitals
with a relatively long range, fine integration grids in real and 
reciprocal space, and a well converged selfconsistency between density 
and potential.
   For the cheap forces we may save on many different parameters,
depending on the system and the properties studied.
   Thus, we may use the local density approximation (LDA),
a minimal single-$\zeta$ basis set with short range, a coarser 
integration grid in real space, just the $\Gamma$ point in reciprocal
space, and the non-selfconsistent Harris
functional~\cite{Harris1985}.
   All together, the cheap forces are typically hundreds of times
faster to compute than the converged ones, and therefore they add
a neglegible cost to the overall calculation, thus making unneccessary
to resort to classical force fields.

   As usual, we use the Born-Oppenheimer approximation and we treat the
nuclei as classical particles, subject to the Hellmann-Feynman forces
(including all Pulay corrections).
   The equations of motion are solved with the standard velocity-Verlet 
algorithm~\cite{Allen-Tildesley1987}, what ensures the time 
reversibility of the trajectories~\cite{Tuckerman1992}.
   The atomic forces at time $t$ are defined as 
${\bf F}_{fast}(t) + \Delta {\bf F}(t)$, where
\begin{equation}
\label{DF}
\Delta {\bf F}(t) = 
   \left\{ 
         \begin{array}{ll}
            n ( {\bf F}_{conv}(t) - {\bf F}_{fast}(t) ) & 
               \mbox{if $ (t/\Delta t \bmod n) = 0$}    \\
            0 & \mbox{otherwise.}
         \end{array}
   \right. 
\end{equation}
   Thus, the expensive converged forces ${\bf F}_{conv}$ need to be 
evaluated only once every $n$ time steps $\Delta t$.
   In those `correction steps', the trajectories generated by the 
cheap (fast) forces ${\bf F}_{fast}$ are corrected by applying a 
force `kick' equal to the difference between the converged and 
fast forces at that time, multiplied by $n$. 
   The factor $n$ accounts for the concentration of the continuous
force correction ${\bf F}_{conv}(t) - {\bf F}_{fast}(t)$ in one
out of every $n$ steps.
   The method of Ref.~\onlinecite{Batcho2001}, based on the 
position-Verlet algorithm, was reported to have a better 
numerical stability in response to the correction `kicks'.
   The efficiency of that method in the present context will be
studied in future works.

\begin{figure}[htbp]
\includegraphics[width=0.8\columnwidth]{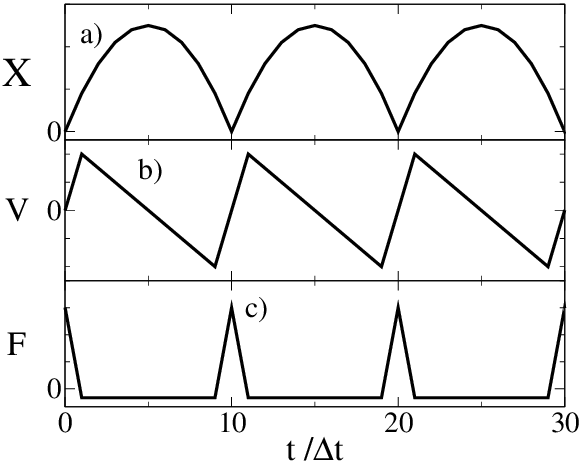}
\caption{
   Schematic a) position, b) velocity, and c) force, in arbitrary units,
for a particle moving in one dimension, generated by the mixed-force
algorithm with zero initial velocity and a force correction interval 
of 10 time steps.
   For simplicity, the converged force is equal to zero and the
cheap (fast) force is a negative constant.
   The periodic force correction `kicks' are positive and invert
the velocity.
   Notice that the position and velocity at the correction steps
are equal to their correct converged value (zero).
}
\label{explanation}
\end{figure}

   Figure~\ref{explanation} shows schematically the positions, 
velocities and forces of a particle moving in one dimension, 
generated with our mixed-force algorithm.
   For simplicity, we take the converged force and the initial 
velocity equal to zero, so that the correct converged position is 
also zero at all times.
   The mixed-force trajectory, for a constant negative fast
force, shows periodic force kicks that change discontinuously the
velocity and invert the trajectory at the correction steps.
   
   We have applied this method to simulate a system of 64 silicon 
atoms at an average temperature of $\sim 2000$~K and an average 
pressure close to zero.
   This high temperature was intentionally chosen to test the
method under specially stringent conditions, whith high kinetic
energies and frequent formation and breaking of bonds.
   The simulations were performed with the SIESTA 
program~\cite{Soler2002} but standard Hamiltonian diagonalizations 
were used instead of order-$N$ 
methods~\cite{Ordejon1993,Kim-Mauri-Galli1995}, 
because of the metallic character of liquid silicon. 
   For the cheap forces we use the Harris functional, 
a minimal basis set with a range of 3.5 and 4.0 Bohr
for $s$ and $p$ orbitals, 
a real-space integration grid with a plane wave cutoff of 40~Ry, 
and only the $\Gamma$ $k$-point.
   For the converged forces, we use the self-consistent Kohn-Sham
functional in the LDA, a double-$\zeta$ polarized (DZP) basis set
with a range of 5.4, 6.5, and 3.8 for $s,p$ and $d$ basis orbitals, 
a real-space grid with a 80~Ry plane wave cutoff, and only the 
$\Gamma$ $k$-point.
   The forces are corrected according to equation~(\ref{DF}) every 
$n$ time steps, with $\Delta t = 1$ fs. 

\begin{figure}[htbp]
\includegraphics[width=0.9\columnwidth]{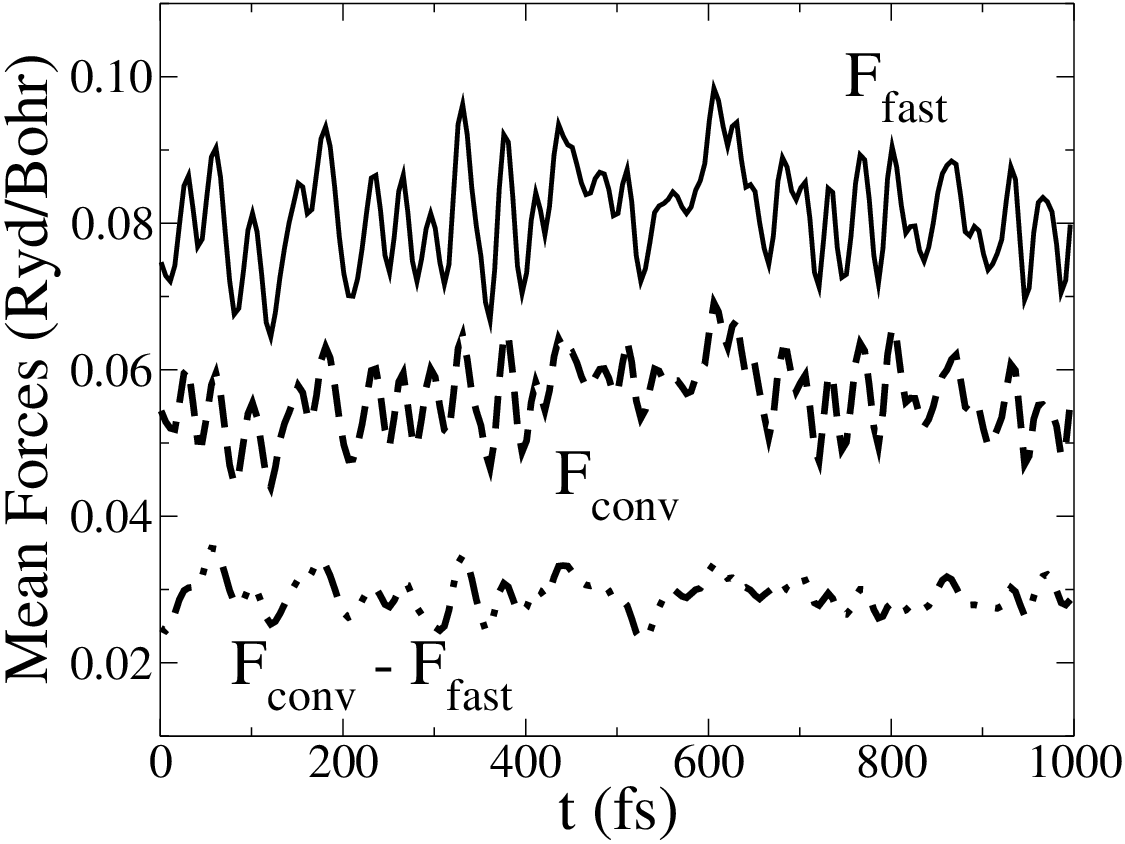}
\caption{
   Decomposition of the total converged forces ${\bf F}_{conv}$
into a cheaply evaluated component ${\bf F}_{fast}$, and a remainder
${\bf F}_{conv} - {\bf F}_{fast}$.
   Represented are the average norms as a function of time. 
   The trajectory was generated for 64 Si atoms at 2000 K, using
the converged forces.
}
\label{forces}
\end{figure}

   Figure~\ref{forces} compares the magnitudes of the fast and
converged forces, and of their difference.
   It can be seen that the latter is a relativelly small and smooth 
correction, what explains why it may be evaluated and applied less 
frequently.

\begin{figure}[!htbp]
\includegraphics[width=0.9\columnwidth]{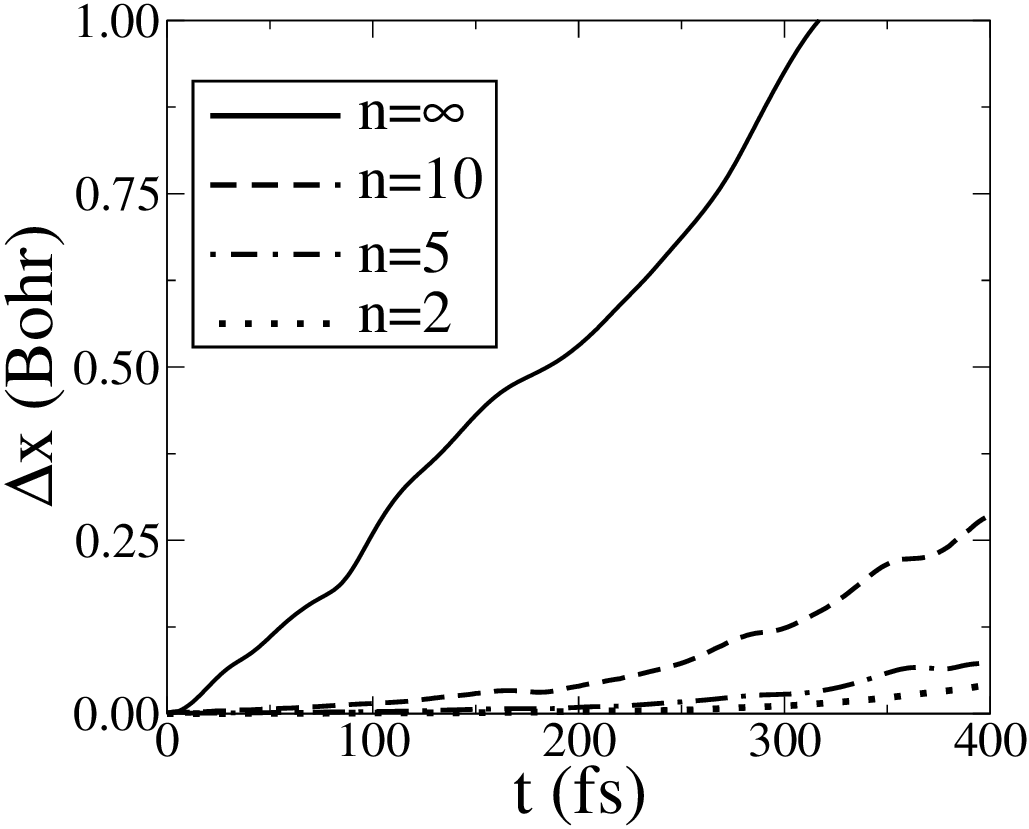}
\caption{
   Divergence of the mixed-force MD trajectories from the 
converged-force trajectory for the liquid silicon system.
   $\Delta x$ is the average distance in atomic positions
between the given and the reference trajectories.
   Force corrections were made every $n=2, 5, 10, 20$, and $\infty$ 
(only fast forces) time steps.
}
\label{divergence}
\end{figure}

   Figure~\ref{divergence} represents the divergence of the 
trajectories, generated with different values of $n$,
away from the reference converged trajectory (which corresponds to
$n=1$ in Eq.~(\ref{DF})).
   It can be seen that even the $n=10$ trajectory diverges much more 
slowly than that obtained purely from the fast forces 
(labelled $n=\infty$, i.e.~with no force corrections).

\begin{figure}[!htbp]
\includegraphics[width=\columnwidth]{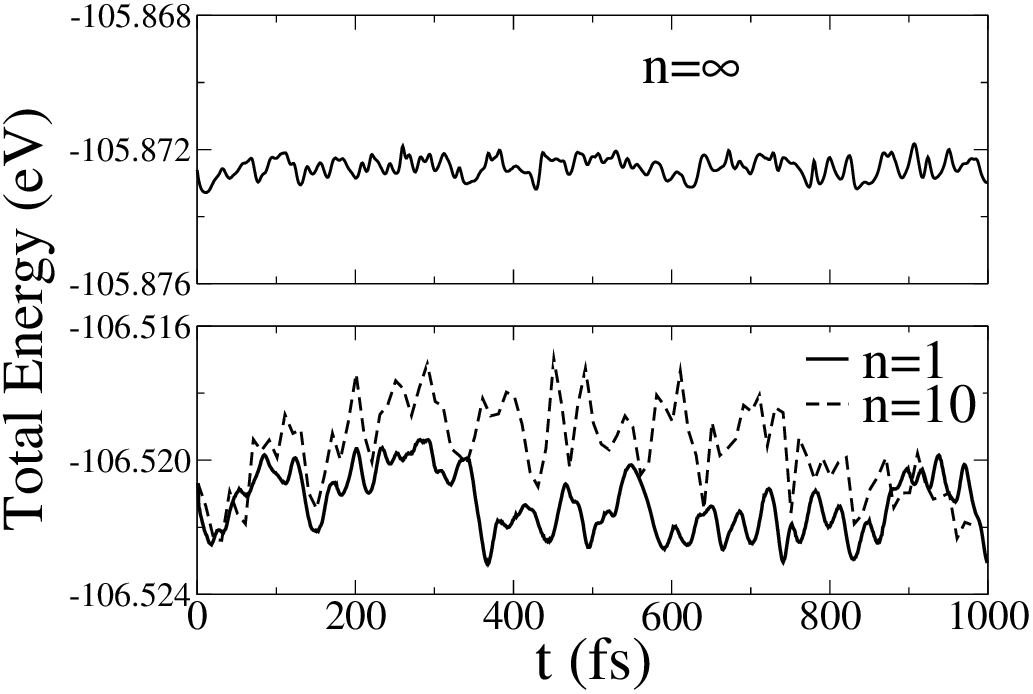}
\caption{
   Total energy per atom, as a function of time, for the liquid 
silicon system.
   In the mixed-force ($n=10$) and converged-force ($n=1$) 
trajectories, the total energy was calculated at the correction 
steps, as the sum of the Kohn-Sham energy plus the nuclear repulsion
and kinetic energies.
   In the fast-force trajectory ($n=\infty$), it was calculated 
at every step, using the Harris-functional for the electronic part.
   The standard deviations are 0.9, 1.3, and 0.3 meV/atom for the
$n=1, 10$, and $\infty$ trajectories, respectively.
}
\label{energy}
\end{figure}

   Figure~\ref{energy} shows the total energy as a function of time.
   The energy conservation is considerably worse in the 
selfconsistent converged-force trajectory ($n=1$) than in the 
Harris-force trajectory ($n=\infty$).
   This probably reflects larger effects of charge 
sloshings and analytic discontinuities in the forces, due to
frequent level crossings in this highly disordered system.
   However, it is important to notice that the energy conservation
in the mixed-force trajectory ($n=10$) is similar to that in the 
converged trajectory.

\begin{figure}[!htbp]
\includegraphics[width=0.9\columnwidth]{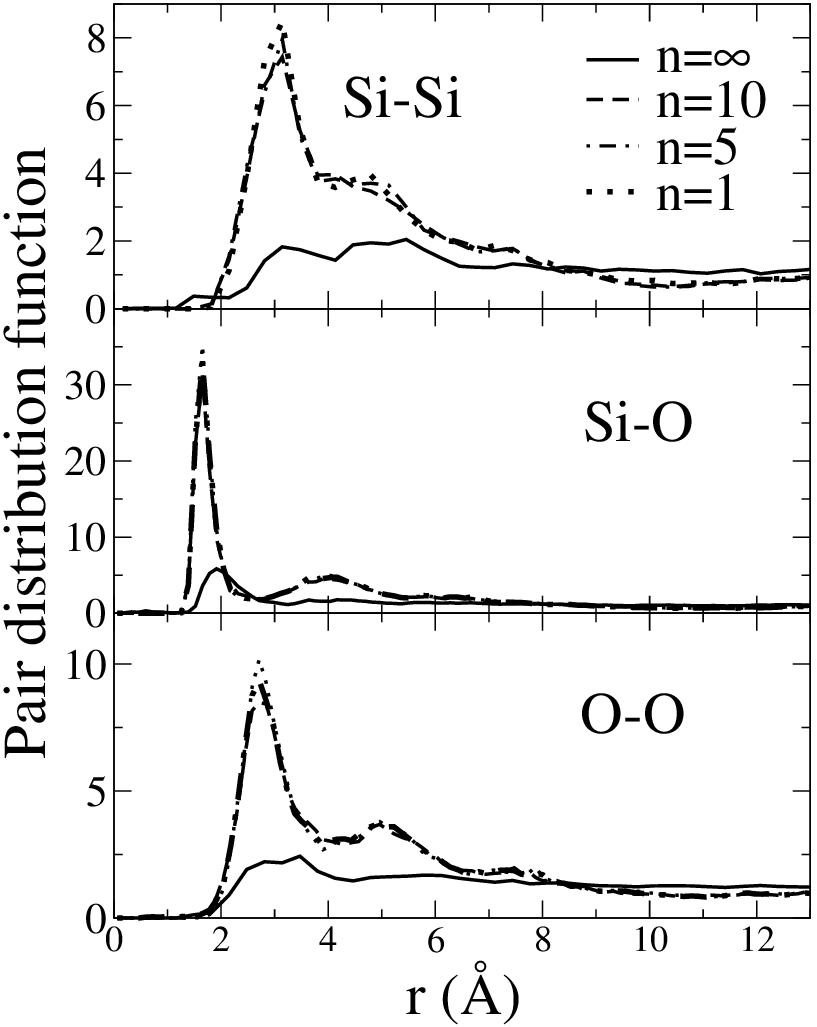}
\caption{
   Radial pair distribution functions of liquid SiO$_2$ at 5500 K, 
using the new method with different values of $n$.
  The converged forces, used for $n=1$ and for the correction steps
of $n=5$ and 10, were obtained  with a double $\zeta$ plus polarization 
basis set, a real-space grid with a plane wave cutoff of 200 Ryd
and only the $\Gamma$ $k$-point. 
   The fast forces (that yield the $n=\infty$ curves when uncorrected) 
were calculated with a minimal basis set (single $\zeta$), a real-space 
grid with a plane wave cutoff of 150 Ryd and only the $\Gamma$ $k$-point.
}
\label{SiO2-distances}
\end{figure}

\begin{figure}[!htbp]
\includegraphics[width=0.9\columnwidth]{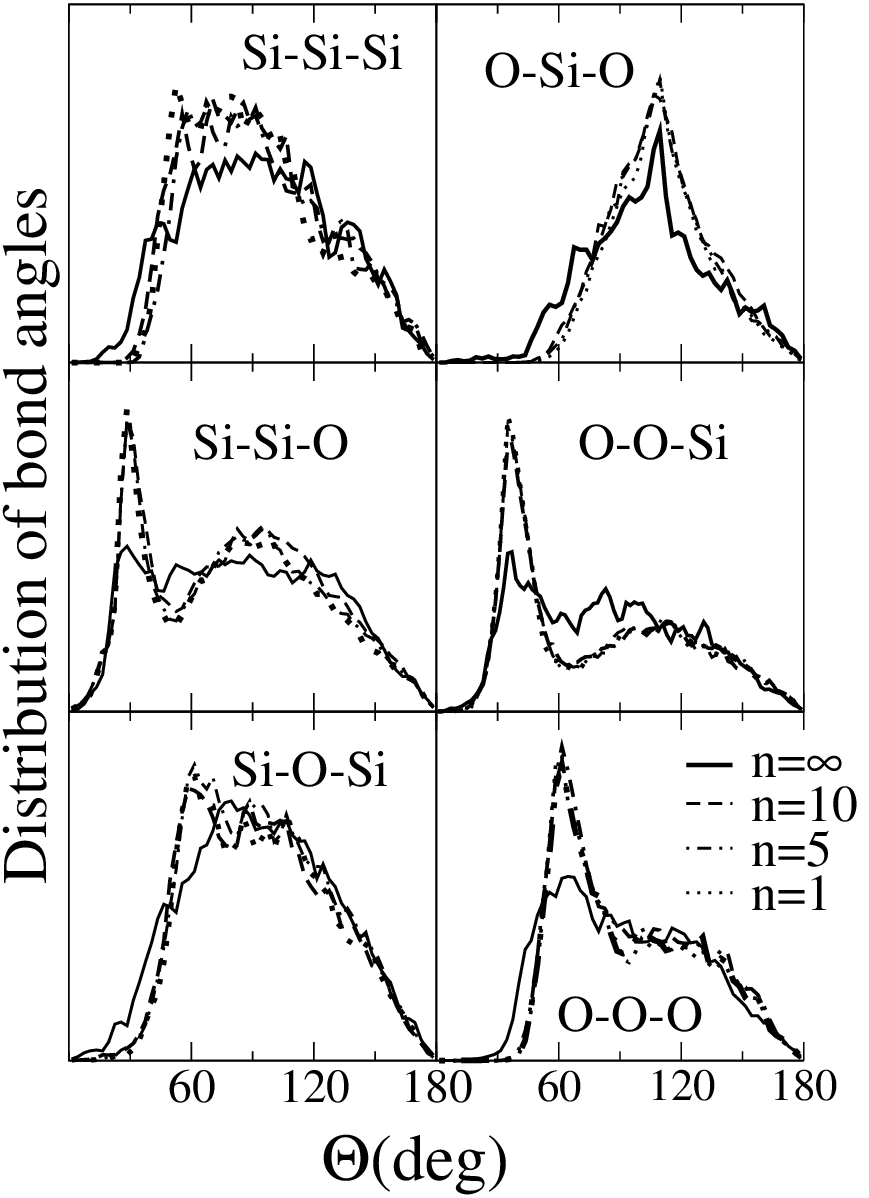}
\caption{
   Bond-angle distribution functions of the liquid SiO$_2$ system, 
using the new method with different values of n.
   The bond cutoff radii were chosen 10\% larger than the first
maximum in the corresponding pair distribution functions of 
Fig.~\ref{SiO2-distances}.
}
\label{SiO2-angles}
\end{figure}

   Despite the high simulation temperature, the charge transfer in 
elemental liquid silicon may be expected to be considerably smaller 
than in an ionic system, making the non-selfconsistent Harris 
functional specially adequate.
   In fact, we have seen that structural magnitudes like the
bond length and bond angle distributions are not very different
using the Kohn-Sham and Harris functionals.
   Therefore, we have also studied a more challenging system,
liquid silica, using 72 atoms at a high average temperature of 5500 K
and a low density of 0.42 g/cm$^3$, typical of porous silica 
aerogels~\cite{Rahmani2001}.
   The distributions of bond legths and angles, presented in 
Figures~\ref{SiO2-distances} and \ref{SiO2-angles}, are indeed 
very different using the two functionals ($n=1$ and $n=\infty$).
   Despite this, the mixed-force method, with up to $n=10$, yields
essentially the same distributions as the converged Kohn-Sham
trajectory. 
   Similar results, to be presented elsewhere, were obtained for
an even more ionic system, liquid magnesium oxide, with 54 atoms 
at 6500 K and 30 GPa.

\begin{figure}[!htbp]
\includegraphics[width=0.9\columnwidth]{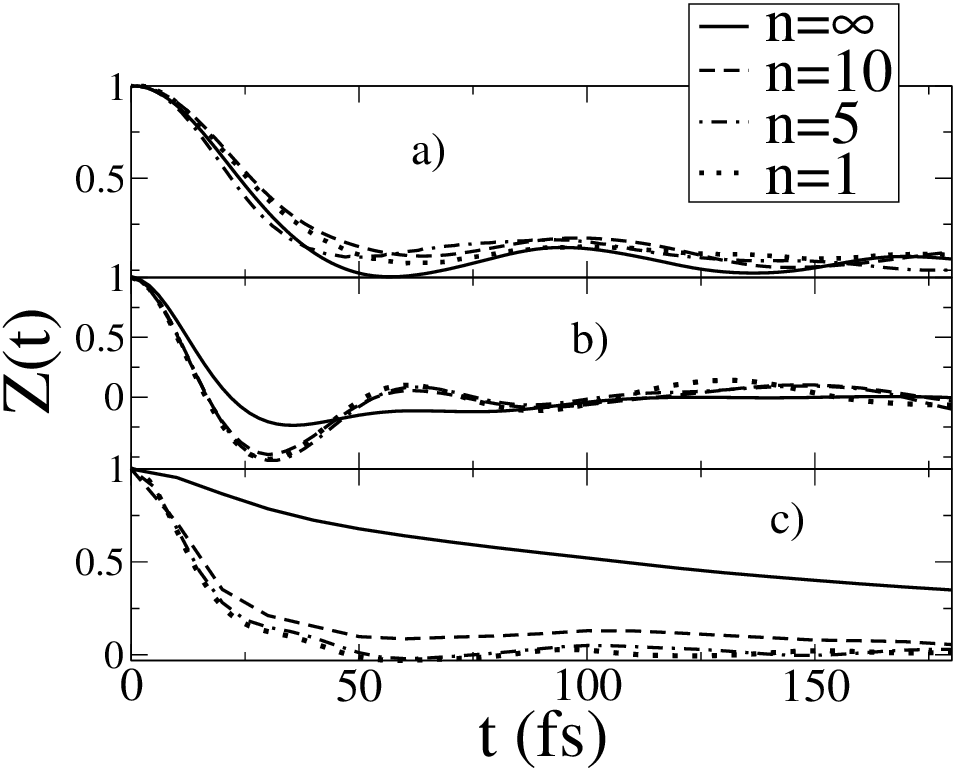}
\caption{
  Velocity autocorrelation function 
$Z(t) = \langle {\bf v}_i(t') {\bf v}_i(t'+t) \rangle /
        \langle {\bf v}_i^2(t') \rangle$,
as a function of the interval $n$ between force corrections, for
a) liquid Si at 2000 K and zero pressure, 
b) liquid MgO at 6500 K and 30 GPa, and 
c) liquid SiO$_2$ at 5500 K and a density of 0.42 g/cm$^3$. 
}
\label{vacf}
\end{figure}

   It might be expected that dynamical magnitudes are 
more sensitive than thermodynamic averages to changes in how the 
MD trajectories are obtained.
   Figure \ref{vacf} shows the velocity autocorrelation function 
for the three systems studied, as a function of the interval $n$
between force corrections.
   As expected, the trajectory of the non-selfconsistent
Harris functional is reasonably accurate only for elemental liquid 
silicon.
   But, in every case, the mixed-force method, with up to $n=10$, 
yields essentially the same velocity autocorrelations as the 
converged Kohn-Sham trajectories.
   We have also calculated self-difussion 
coefficients from the average quadratic distances traversed as
a funtion of time. 
   Thus, for liquid silicon we obtain, respectively, 
$(2.4 \pm 0.1)$, 
$(2.5 \pm 0.1)$,
$(2.6 \pm 0.1)$, 
$(2.6 \pm 0.1)$, and 
$(2.0 \pm 0.1)\times10^{-4}$  $\textrm{cm}^2/\textrm{s}$, 
for $n=1, 5, 10, 20$ and $\infty$.
   Again, the mixed-force value, even with $n=20$, is the same,
within the statistical error, as that of the converged trajectory.
   This indicates that dynamical and kinetic magnitudes, as well as
structural or thermodynamic averages, are well reproduced even with
quite large values of the boost factor $n$.

   In conclusion, we have presented a new method to greatly accelerate
ab initio molecular dynamics simulations by combining cheap force 
evaluations with accurate converged ones.
   Our results show that the method is very robust with respect to 
the reduced accuracy of the cheap forces.
   Although the acceleration factor will undoubtly depend on the
system simulated, our present results indicate that factors of 10
can be expected in most cases.

   We thank specially Gabriel Fabricius for discussions, for help 
with the liquid silicon parameterization, and for sharing with us
his data-processing programs.
   We also acknowledge useful discussions with Emilio Artacho.
   This work has been supported by the Fundaci\'on Ram\'on Areces and
by Spain's MCyT grant BFM2000-1312.

\bibliographystyle{apsrev}
\bibliography{dft,siesta,misc,fmix}

\end{document}